\documentclass[12pt]{iopart}
\usepackage{epsfig}
\begin{document}
\jl{1}
\newfont{\fraktur}{eufm10 at 11pt}

\letter{Nonabelian density functional theory}
\author{G Rosensteel and Ts Dankova}
\address{Physics Department, Tulane University, New Orleans LA 70118 USA}

\begin{abstract}
Given a vector space of microscopic quantum observables, density functional theory is formulated on its dual space. A generalized Hohenberg-Kohn theorem and the existence of the universal energy functional in the dual space are proven. In this context ordinary density functional theory corresponds to the space of one-body multiplication operators. When the operators close under commutation to form a Lie algebra, the energy functional defines a Hamiltonian dynamical system on the coadjoint orbits in the algebra's dual space. The enhanced density functional theory provides a new method for deriving the group theoretic Hamiltonian on the coadjoint orbits from the exact microscopic Hamiltonian.
\end{abstract}

\pacs{03.65.Fd, 21.60.Fw, 31.15.Ew, 71.15.Mb}
\submitted

\maketitle

\section{INTRODUCTION}

Density functional theory, in its simplest form, is a theory for nondegenerate ground states that replaces the many particle antisymmetrized wave function by its density $\rho(\vec{r})$. The Hohenberg-Kohn theorem shows that there exists a universal energy functional minimized by the density of the exact ground state wave function \cite{HK64,Lieb83}. In addition to ubiquitous applications in quantum chemistry \cite{Paar}, this many body theory solves problems in solid state \cite{Callaway,Langreth} and nuclear physics \cite{Dreizler}.

This letter's aim is to generalize density functional theory to a framework where the densities are elements of the dual space to a vector space of microscopic observables. The Hohenberg-Kohn theorem is proven in this context, and the existence of the universal energy functional on the dual space is established. If the vector space is also a Lie algebra, then a group theoretic model is defined naturally on the coadjoint orbits of the algebra's dual space. The generalized Hohenberg-Kohn theorem enables the derivation of the group theoretic Hamiltonian from the microscopic Hamiltonian on Fock space. The problem of constructing an algebraic model Hamiltonian from the microscopic interaction is an old one in group theory with many potential applications to the description of complex many-body systems. 

Let ${\cal H}$ denote the Hilbert space of $N$-fermion antisymmetrized state vectors, ${\cal P}({\cal H})$ its complex projective space, and $H$ the self-adjoint Hamiltonian operator on ${\cal H}$. Suppose {\fraktur g} is a real linear vector space of hermitian operators defined on ${\cal H}$, and {\fraktur g}$^{\ast}$ is its dual space, i.e., the vector space of real-valued linear functionals on {\fraktur g}. Denote the pairing between a dual element $\rho \in \mbox{\fraktur g}^{*}$ and an operator $X \in \mbox{\fraktur g}$ by $\langle \rho , X \rangle \equiv \rho(X)$. Define the moment map $M : \cal{P}(\cal{H})$ $\rightarrow \mbox{\fraktur g}^{*}$ from the projective space of states to the dual space by 
\begin{equation}
\langle M(\phi), X \rangle := \frac{\langle \phi | X \phi \rangle}{\langle \phi | \phi \rangle} ,
\end{equation}
where $\phi \in {\cal H}$, $X\in \mbox{\fraktur g}$, $M(\phi)\in \mbox{\fraktur g}^{*}$. A brief history of the moment map is given on p.\,327 of \cite{Marsden}.
Define the energy mapping $F: {\cal P}({\cal H})\rightarrow {\bf R}$ from the projective space to the real numbers by
\begin{equation}
F(\phi) := \frac{\langle \phi | H \phi \rangle}{\langle \phi | \phi \rangle} ,
\end{equation}
and for each fixed $X\in \mbox{\fraktur g}$, define $F_{X} : {\cal P}({\cal H})\rightarrow {\bf R}$ by
\begin{equation}
F_{X}(\phi) := F(\phi) + \langle M(\phi) , X \rangle .
\end{equation}

For most physical systems only a few quantum states are observed or are of direct interest, e.g., the ground state, rotational and vibrational bands, collective states, and resonances. Although the microscopic theory on the Hilbert space ${\cal H}$ is complete, it may obfuscate the true physics because all degrees of freedom are treated democratically. By selecting out the most relevant degrees of freedom, a discriminating  model based on the observables in {\fraktur g} can elucidate the physics and simplify the mathematics. The moment  mapping is an indispensable tool for creating a tractable theoretical description of microscopic many-body systems. The density $\rho = M( \phi )$ retains only part of the entire information about the system that the wave function $\phi$ carries, but a very important part -- the expectations of the observables that span {\fraktur g}.

When {\fraktur g} is a subspace of real-valued measurable functions $v(\vec{r})$ on {\bf R}$^3$, and {\fraktur g}$^{\ast}$ is the space of ${\cal L}^{1}$ functions
\begin{equation}
\mbox{\fraktur g}^{*} =  \left\{ \rho : {\bf R}^{3}\rightarrow {\bf R}\  \left| \  \int | \rho(\vec{r}) | \, d^{3}r < \infty \right\} , \right.
\end{equation}
conventional density functional theory is recovered. The elements of {\fraktur g} are regarded as hermitian multiplication operators on ${\cal H}$ by associating the one-body operator $\sum_{n=1}^{N} v(\vec{r}_{n})$ with the function $v(\vec{r})\in \mbox{\fraktur g}$. The pairing is given by $\langle \rho , v \rangle = \int \rho(\vec{r}) v(\vec{r}) d^{3}r$. If $\phi$ is normalized, then the moment map $M(\phi)=\rho$ is
\begin{equation}
\rho(\vec{r}) =  N \int | \phi(\vec{r}, \vec{r}_{2},\ldots ,\vec{r}_{N}) |^2 \, d^{3}r_{2}\ldots d^{3}r_{N}.
\end{equation}
The functional $F_{v}(\phi)$ is the expectation of the energy including the one-body external potential $\sum_n v(\vec{r}_{n})$. 

Von Neumann's density matrix formulation of quantum mechanics is included in the formalism presented here by letting {\fraktur g} be the space of all (not just one-body) hermitian operators on ${\cal H}$. The dual space {\fraktur g}$^{\ast}$ may be identified with a space of hermitian operators where the pairing is defined by $\langle \rho, X \rangle = {\mbox tr} ( \rho X )$. The moment map is given by $M(\phi) = |\, \phi \, \rangle \langle \, \phi \, |$ if $\phi \in {\cal H}$ is normalized. In this case, the moment map is one-to-one, and the microscopic theory is physically and mathematically equivalent to the density matrix expression of quantum mechanics. 

Another example is provided by cranking Hamiltonians or Routhians for which {\fraktur g} and $\mbox{\fraktur g}^{\ast}$ are isomorphic to the real three-dimensional vector space {\bf R}$^{3}$. The hermitian operator corresponding to $\vec{\omega}\in \mbox{\fraktur g}$ is $\vec{\omega} \cdot \vec{J}$, where $\vec{J}$ denotes the one-body angular momentum operators on ${\cal H}$. The  pairing is the dot product, $\langle  \vec{\rho} , \vec{\omega} \rangle = \vec{\rho}\cdot \vec{\omega}$, and the moment map is $M(\phi) = \vec{\rho}$, where 
\begin{equation}
\rho_{k} =  \frac{\langle \phi | J_{k} \phi \rangle}{\langle \phi | \phi \rangle} . 
\end{equation}
Note that $F_{\omega}(\phi)$ is the expectation of the cranking Hamiltonian.

In the case of the cranked Nilsson model, the space {\fraktur g} consists of all hermitian operators formed from linear combinations of the angular momentum $\vec{J}$ and the quadrupole operator $Q^{(2)}_{\mu} = \sum_{n} r^2_{n} Y^{(2)}_{\mu}(\Omega_{n})$. This eight dimensional space {\fraktur g} is the rotational model Lie algebra $rot(3)$. The full Nilsson model expands {\fraktur g} to a 10 dimensional space that includes the spin-orbit interaction and the square of the orbital angular momentum. Several more examples are produced by taking the vector space of operators to be the Elliott $su(3)$ algebra \cite{Elliott}, the Ginocchio $so(8)$ fermion algebra \cite{Ginocchio}, the general collective model algebra $gcm(3)$ \cite{Rose76}, the symplectic model algebra $sp(3,{\bf R})$ \cite{Rose77}, and the unitary algebra of all one-body operators. In many of these examples the vector space {\fraktur g} is a Lie algebra with respect to operator commutation.  In \S 3 it is shown that when this additional algebraic structure is present, a mean field theory is naturally defined on the coadjoint orbit surfaces in the dual space {\fraktur g}$^{\ast}$ .

\section{HOHENBERG-KOHN-LEVY}
In this section the fundamental Hohenberg-Kohn theorem is proven in the extended framework of an arbitrary vector space of operators {\fraktur g} on Fock space. The Levy constrained-search formulation is also established in this general case.

Define the set of states ${\cal S}$ consisting of every $\phi \in \cal P(H)$ that is an absolute nondegenerate minimum of $F_{X}$ for some $X \in \mbox{\fraktur g}$. Set ${\cal C}$ equal to the image of ${\cal S}$ under the moment map, ${\cal C} = M({\cal S}) \subset \mbox{\fraktur g}^{*}$. 

\vspace{2mm}

\noindent {\bf Theorem 1}. The moment map $M : {\cal S} \rightarrow {\cal C}$ is an injection.
\newline  {\it Proof:} The proof of this theorem is similar to the original Hohenberg-Kohn argument \cite{HK64}. Suppose, to the contrary, that there are two distinct states $\phi_{1}, \phi_{2}\in {\cal S}$, $\phi_{1}\neq \phi_{2}$, yet $M(\phi_{1})=M(\phi_{2})$. There are operators $X_{1}, X_{2}\in \mbox{\fraktur g}$ such that $\phi_{1}$ is an absolute nondegenerate minimum for $F_{X_{1}}$ and $\phi_{2}$ is an absolute nondegenerate minimum for $F_{X_{2}}$. Since $\phi_{1}\neq \phi_{2}$, the nondegeneracy of the minima implies
\begin{eqnarray*}
E_{1} \equiv F_{X_{1}}(\phi_{1}) & < & F_{X_{1}}(\phi_{2}) \\
E_{2} \equiv F_{X_{2}}(\phi_{2}) & < & F_{X_{2}}(\phi_{1}) .
\end{eqnarray*}
{\fraktur g} is a vector space, so $X_{1}-X_{2} \in \mbox{\fraktur g}$ and
\begin{eqnarray*}
E_{1} & < & F_{X_{1}}(\phi_{2}) = F(\phi_{2}) + \langle M(\phi_{2}), (X_{1}-X_{2}) + X_{2} \rangle \\
& = & F_{X_{2}}(\phi_{2}) + \langle M(\phi_{2}), (X_{1}-X_{2}) \rangle \\
{\mbox or \ } E_{1} & <  & E_{2} + \langle M(\phi_{2}), (X_{1}-X_{2}) \rangle . 
\end{eqnarray*}
By repeating the argument for interchanged $1$ and $2$,
\[ E_{2} < E_{1} + \langle M(\phi_{1}), (X_{2}-X_{1}) \rangle . \]
Adding the two equations, and recalling that $M(\phi_{1})=M(\phi_{2})$, leads to the contradiction
\[ E_{1} + E_{2} < E_{2} + E_{1} . \]
Hence, for two distinct states in ${\cal S}$ the moment map yields two distinct densities in ${\cal C}$.

\vspace{2mm}

The theorem depends only on the vector space structure of {\fraktur g}, not on any possible additional algebraic structure. It is a corollary that for each $X \in \mbox{\fraktur g}$, the mapping $\tilde{F}_{X} = F_{X} \circ M^{-1} :$
${\cal C}$ $\rightarrow {\bf R}$ is well-defined. The state $\phi$ minimizes $F_{X}$ on $S$ if and only if $\rho = M(\phi)$ minimizes $\tilde{F}_{X}$ on ${\cal C}$.

A density $\rho$ is said to be $v$- or, in the general setting, $X$-representable if $\rho\in {\cal C}=M({\cal S})$. The energy functional must be minimized on the set of $X$-representable densities. If the $X$-representable densities can only be characterized by their defining property as an image of a set of $N$-fermion wave functions, this minimization is equivalent to the original many-body eigenvalue problem on Fock space. Another problem is that the existence theorem's proof does not provide a construction of the energy functional.

These two problems are circumvented by the constrained search formulation of Levy \cite{Levy79,Levy82}. Consider the image of the moment map ${\cal C}_{N}=M(\cal{P}(\cal{H}))$, for which $\mbox{\fraktur g}^{\ast}\supset {\cal C}_{N}  \supset {\cal C}$. A density $\rho$ is said to be $N$-representable if $\rho\in {\cal C}_{N}$. Define the energy functional
\begin{eqnarray}
E & : & {\cal C}_{N} \rightarrow {\bf R} \nonumber \\
E[\rho] & = & \inf_{\phi \in M^{-1}(\rho)} F(\phi) , \label{efunctional}
\end{eqnarray}
and also $E_{X}[\rho] = E[\rho] + \langle \rho , X\rangle $ for $\rho\in {\cal C}_{N}$.

\vspace{2mm}

\noindent {\bf Theorem 2}. The ground state energy $E_{0}$ satisfies the minimization condition
\[ E_{0} = \inf_{\rho \in {\cal C}_{N}} E_{X}[\rho] , \]
and the ground state density $\rho_{0}$ attains this minimized energy. If $\rho$ is $X$-representable, then $E_{X}[\rho] = \tilde{F}_{X}(\rho)$.
\newline {\it Proof:} The ground state energy minimizes the expectation of the total Hamiltonian:
\begin{eqnarray*}
E_{0} & = & \min_{\phi\in {\cal H}} F_{X}(\phi)  \\
& = & \min_{\rho \in {\cal C}_{N}} \left\{ \min_{\phi\in M^{-1}(\rho)}  ( F(\phi) + \langle M(\phi) , X\rangle  )  \right\} \\
& = & \min_{\rho \in {\cal C}_{N}} (E[\rho] + \langle \rho, X\rangle ) .
\end{eqnarray*}
If $\phi_{0}\in {\cal S}$ is a nondegenerate ground state, $\rho_{0}\in {\cal C}$ its density, and $E_{0} = E_{X}[\rho_{0}]$ the ground state energy, then
\[ E_{0} = F_{X}(\phi_{0}) = \tilde{F}_{X}(\rho_{0}). \]

\vspace{2mm}

Thus, according to the generalized Levy's theorem, it is sufficient to minimize the energy functional on the space of all $N$-representable densities instead of all $X$-representable densities.

To determine the universal energy functional $E[\rho]$, the Lagrange multiplier theorem, suitably extended to Banach spaces, may be applied to eliminate the constraint to $\phi \in M^{-1}(\rho)$ from the functional's defining equation (\ref{efunctional}). This is achieved by regarding the pairing $\langle \rho , X\rangle $ as a Lagrange multiplier term while the operator $X$ is the Lagrange multiplier itself. Suppose the vector space of operators is self-dual, i.e., the dual space to $\mbox{\fraktur g}^{\ast}$ is {\fraktur g}, and it is a Banach space, i.e., a norm $\|X\|$ is defined on {\fraktur g}. Both conditions are satisfied, for example, if the space is finite-dimensional. Suppose that $M^{-1}(\rho)$ is a submanifold of ${\cal P}({\cal H})$.

\vspace{2mm}

\noindent {\bf Theorem 3}. The following are equivalent conditions on $\phi_{0} \in M^{-1}(\rho)$:
\newline (i) $\phi_{0}$ is a critical point of $F(\phi)$ restricted to $M^{-1}(\rho)$; and
\newline (ii) there is a $X_{0}\in \mbox{\fraktur g}$ such that $(X_{0},\phi_{0})$ is a critical point of $F_{X}(\phi)$ on the space $\mbox{\fraktur g} \times {\cal P}({\cal H})$.

\vspace{2mm}

\noindent The universal energy functional is given by $E[\rho]=F(\phi_{0})$ when the critical point corresponds to an absolute minimum. To test for a minimum, let $\alpha$ be a real positive constant, $(X_{0},\phi_{0})$ a critical point of $F_{X}(\phi)$, and set
\begin{equation}
F^{\alpha}_{X}(\phi) = F_{X}(\phi)  + \alpha \| X-X_{0} \|^{2},
\end{equation}
that also has a critical point at $(X_{0},\phi_{0})$. If the restriction of $F(\phi)$ to $M^{-1}(\rho)$ has a nondegenerate minimum at $\phi_{0}$, then for sufficiently large positive $\alpha$, $F^{\alpha}_{X}(\phi)$ is minimized at $(X_{0},\phi_{0})$. This observation may be used to identify minima using the unconstrained second derivative test. The Newton method, in its Banach space generalization, may be applied to compute critical points for $F^{\alpha}_{X}(\phi)$ \cite{Rowe77}. See p.\,219 of \cite{Marsden} for the theorem's proof.

\section{COADJOINT ORBITS}
Suppose the vector space of hermitian operators {\fraktur g} is a Lie algebra and let $G$ denote the corresponding simply connected and connected Lie group of unitary operators. The group action of $G$ on Fock space naturally induces the coadjoint action $Ad_{g}^{*}$ of $G$ on the dual space of densities. To show this, suppose $\phi \in {\cal H}$ and $g\cdot \phi \in {\cal H}$ is the result of the unitary representation of $g\in G$ acting on $\phi$. If $\rho = M(\phi)$, then $M(g\cdot \phi) = Ad_{g}^{*}(\rho)$, since
\begin{eqnarray}
\langle M(g\cdot \phi), X \rangle & = & \langle g\cdot \phi | X\, g\cdot \phi \rangle /  \langle  g\cdot \phi |  g\cdot \phi \rangle \nonumber \\
& = &  \langle \phi | g^{-1} X g \cdot \phi \rangle  /  \langle \phi |  \phi \rangle \nonumber \\
& = & \langle \rho, Ad_{g^{-1}} (X)  \rangle \nonumber \\
& = & \langle Ad_{g}^{*}(\rho) , X  \rangle 
\end{eqnarray}
for all $X\in \mbox{\fraktur g}$. Thus the coadjoint action leaves the space of $N$-representable densities invariant, $Ad_{g}^{\ast} : {\cal C}_{N} \rightarrow {\cal C}_{N}$.

If $G$ is a symmetry group of the Hamiltonian $H$, then the energy functional transforms according to the adjoint transformation,
\begin{equation}
E_{X} \circ Ad_{g}^{\ast} = E_{Ad_{g^{-1}} (X)},
\end{equation}
for $X\in \mbox{\fraktur g}$ and $g\in G$. If $\rho \in {\cal C}$ is an absolute nondegenerate minimum for $E_{X}$, then $Ad_{g}^{*}(\rho)$ is an absolute nondegenerate minimum for $E_{Ad_{g^{-1}} (X)}$. Therefore, if $G$ is a symmetry group for the Hamiltonian, the coadjoint $G$-action leaves the space of $X$-representable densities ${\cal C}$ invariant.

In conventional density functional theory, the abelian Lie algebra integrates to an abelian group of unitary operators
\begin{equation}
G = \left\{ g = \exp \left( i\, \sum_{n=1}^{N} f(\vec{r}_{n}) \right)  ,  f \in \mbox{\fraktur g}  \right\} ,
\end{equation}
and G is the isotropy subgroup at each density, $Ad_{g}^{*}(\rho) = \rho$. If $\phi\in {\cal H}$ and $M(\phi)=\rho \in {\cal C}_{N}$, then the energy functional satisfies
\begin{equation}
E[\rho] \leq \inf_{f\in \mbox{\fraktur g}} F( g\cdot \phi ) .
\end{equation}
This equation provides an upper bound on the exact energy functional and may be used to approximate it.

For a nonabelian Lie algebra {\fraktur g}, the Lie group $G$ is not typically contained in the isotropy subgroup $G_{\rho}$ at $\rho \in {\cal C}_{N}$. The coadjoint orbit ${\cal O}_{\rho}$ consists of all points $Ad_{g}^{\ast}(\rho) \in {\cal C}_{N}$ as $g$ ranges over the group $G$. Each orbit space ${\cal O}_{\rho}$ is diffeomorphic to the homogeneous coset space $G/G_{\rho}$. Hence the space of $N$-representable densities is a union of disjoint coadjoint orbits each diffeomorphic to a homogeneous space.

Every Lie algebra element $X$ defines a tangent vector to the coadjoint orbit surface through the $N$-representable density $\rho$. This vector is the tangent to the curve $\gamma_{X}(t) = Ad^{\ast}_{g}(\rho)$ where $g = \exp (i\,t\,X)$, see Figure. Note that the elements of the isotropy subalgebra $\mbox{\fraktur g}_{\rho}$ correspond to null tangent vectors, and the tangent space is isomorphic to the coset vector space $\mbox{\fraktur g}/\mbox{\fraktur g}_{\rho}$. For any two tangent vectors $X,Y\in \mbox{\fraktur g}/\mbox{\fraktur g}_{\rho}$ to the coadjoint orbit through $\rho$, the bilinear form
\begin{equation}
\omega_{\rho}(X,Y) = i \, \langle  \rho, [X,Y] \rangle  , 
\end{equation}
is antisymmetric, closed, and nondegenerate. Thus each coadjoint orbit is a symplectic manifold \cite{Kirillov}. Moreover the group $G$ acts on each orbit as a transitive group of canonical transformations,
\begin{equation}
\omega_{Ad^{\ast}_{g}(\rho)}(Ad_{g}(X), Ad_{g}(Y)) = \omega_{\rho}(X,Y), 
\end{equation}  
where $X, Y$ are tangent vectors at $\rho$ and $Ad_{g}(X)$ and $Ad_{g}(Y)$ are tangent vectors at $Ad^{\ast}_{g}(\rho)$.

The model group theoretic Hamiltonian $h$ may be defined now. Choose a  coadjoint orbit ${\cal O}_{\rho}$ in the dual space and let ${\cal E}$ denote the restriction of the energy functional to this orbit.  If $X$ is a tangent vector at $\rho$, denote the derivative of the energy in the $X$ direction by
\begin{equation}
d{\cal E}_{\rho} (X) =  \frac{d}{dt} {\cal E}(Ad_{g}^{\ast} (\rho) ) |_{t=0} ,\ \mbox{where\ } g = \exp(i\,t\,X) .
\end{equation}
The model Hamiltonian $h\in \mbox{\fraktur g}/\mbox{\fraktur g}_{\rho}$ is determined by
\begin{equation}
d{\cal E}_{\rho} (X) = \omega_{\rho}(h,X) \mbox{\ for all $X\in \mbox{\fraktur g}$} . \label{omega}
\end{equation}
There is a unique solution to this equation for $h\in \mbox{\fraktur g}/\mbox{\fraktur g}_{\rho}$ because the symplectic form $\omega$ is nondegenerate on the tangent space. Note that the model Hamiltonian $h$ depends on the point $\rho$ in the coadjoint orbit.

In the special case when the Hamiltonian $H$ is an element of the Lie algebra {\fraktur g}, the energy functional is just $E[\rho]=\langle \rho,H\rangle $ and the model Hamiltonian $h$ is equal to $H$. Otherwise the model Hamiltonian $h$ is the optimal approximation to $H$ relative to the Lie algebra {\fraktur g} in the sense that $\langle \phi \, | \, [H-h, X]\, \phi\rangle  = 0$ for all $X \in \mbox{\fraktur g}$ where $M(\phi)=\rho \in {\cal C}$.

If {\fraktur g} is the algebra of one-body operators and the coadjoint orbit consists of the determinental densities, then the energy ${\cal E}$ is inherited uniquely from Fock space because of the one-to-one correspondence between determinental wave functions and determinental densities. The model Hamiltonian $h$ on the manifold of determinental densities is identical to the Hartree-Fock mean field \cite{Rose81a}. But the mean field Hamiltonian $h$ is also defined here for all orbits, including those for the nondetermental densities. The required energy on the general orbits is the universal energy functional $E[\rho]$.  Hartree-Fock-Bogoliubov is extended similarly by this construction using the algebra of quadratic quasiparticle operators and considering the energy functional on generic orbits of quasiparticle densities \cite{Rose81b}.

\section{CONCLUSION}
A useful theoretical description of a many-body system is based on a wise choice for the operator space {\fraktur g}. If the  degrees of freedom that are most  relevant to the dynamics are not represented in {\fraktur g}, then the energy functional $E[\rho]$ on the dual space will be intractable.  The external Coulomb potential created by the positive charges of the massive nuclei dominates the physics of atomic, molecular, and solid state systems. The relatively weak interaction among the electrons can be treated in perturbation theory.  Thus ordinary density functional theory, for which {\fraktur g} is the algebra of one-body multiplication operators, is the appropriate theory for these interacting fermion systems. Since the algebra of multiplication operators is abelian, the coadjoint orbits are just points and no mean field approximation may be constructed in this case.

There is no external potential acting on an atomic nucleus, and the dynamics is determined by a competition among short range pairing, long-range quadrupole-quadrupole, spin-orbit, and other hadronic internucleon forces. The relative influence of the different forces varies with the isotope. Hence there is no obvious choice for {\fraktur g} that applies to all nuclei. If single-particle degrees of freedom are important, then the space of operators should be the unitary algebra $u({\cal H})$ of all one-body fermion operators, i.e., the span of the operators $b^{\dagger}_{\beta} b_{\alpha}$ that destroy a fermion in state $\alpha$ and replace it with one in state $\beta$. If geometrical collective modes are active, then the symplectic algebra $sp(3,{\bf R})$ or one of its subalgebras is suggested \cite{Rose77}. Since the unitary algebra $u({\cal H})$ and the symplectic algebra $sp(3,{\bf R})$ are not abelian, a mean field approximation may be applied to these Lie algebras {\fraktur g}.

\ack G R thanks Jerry Goldstein, Mel Levy and John Perdew for inspiring this work.

\newpage
\begin{figure}[htbp]\centering
\mbox{\psfig{figure=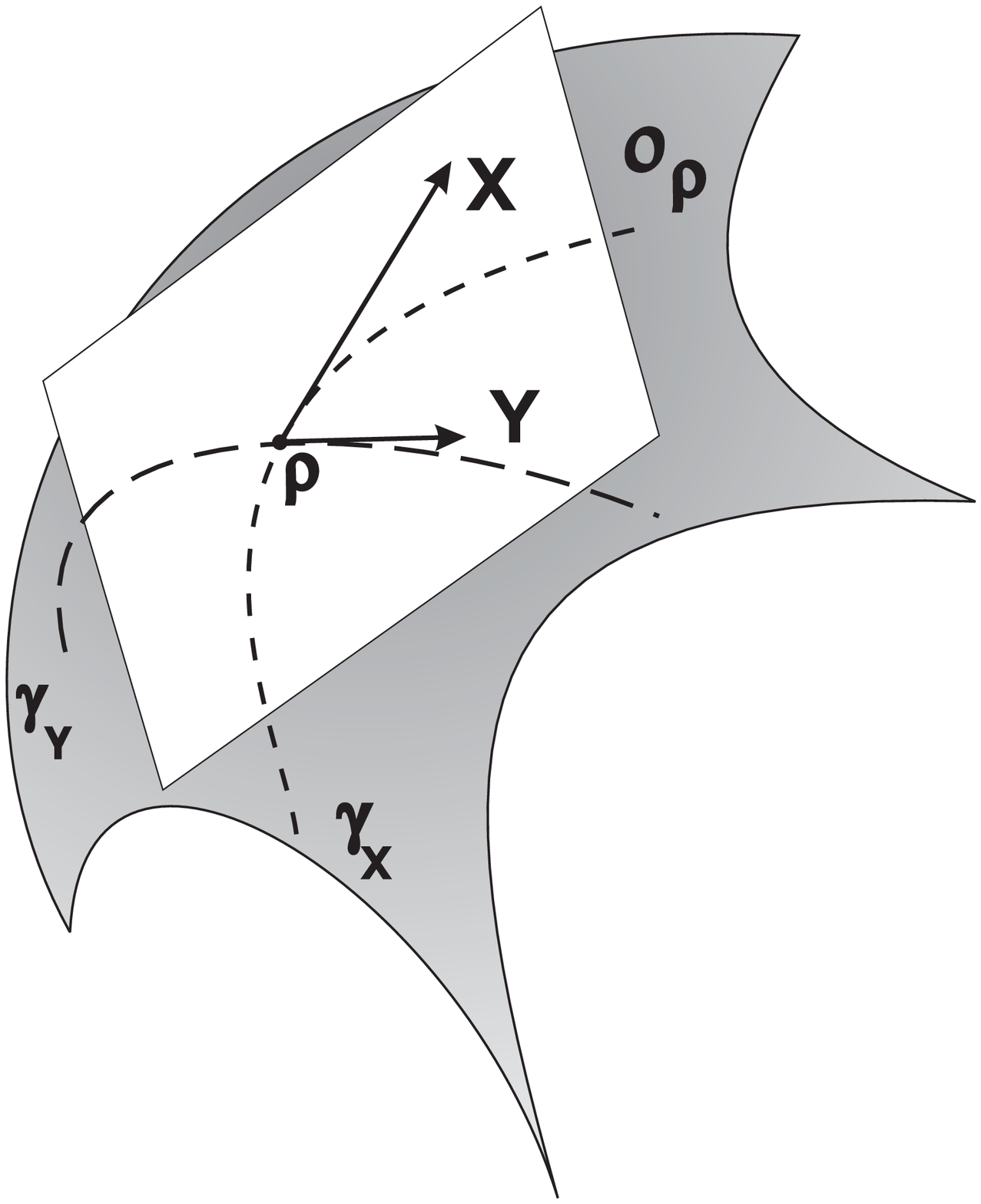,height=4in,width=4in,clip=}}
\caption{The Lie algebra elements $X$ and $Y$ are geometrically viewed as tangent vectors to the curves $\gamma_{X}$ and $\gamma_{Y}$ in the coadjoint orbit surface ${\cal O}_{\rho}$.}
\end{figure}

\end{document}